\documentstyle[preprint,aps]{revtex}

\begin{document}
\draft

\title{ Theoretical Investigation of C$_{\rm 60}$ Infrared Spectrum}
\author{Jaroslav Fabian}
\address{Department of Physics, SUNY at Stony Brook, Stony Brook, NY
 11794-3800}
\maketitle

\vspace{1cm}

\begin{abstract}
A semi-empirical model of the infrared (IR) spectrum of  the C$_{60}$ 
molecule is proposed.  The weak IR-active modes seen experimentally
in a C$_{60}$ crystalline sample are argued to be combination
modes caused by anharmonicity.
The origin of these 2-mode excitations can be either mechanical 
(anharmonic interatomic forces) or  electrical
(nonlinear dipole-moment expansion in normal modes coordinates). 
 It is shown that the 
electrical anharmonicity model exhibits basic features of the 
experimental 
spectrum while nonlinear dynamics would lead to a qualitatively 
different overall picture.
\end{abstract}
\vspace{2em}
\pacs{PACS numbers: 33.20.Ea, 33.20.Tp, 33.70.-w,78.30.-j}

\narrowtext
\newpage

\section{Introduction}
\label{sec:intr}

There has been a great deal of progress in our understanding of the
chemistry and physical properties of fullerenes. Discovery of 
superconductivity
in alkali-metal doped C$_{60}$ \cite{hebard} has ignited discussions 
on possible mechanisms of this phenomenon \cite{schluter,kivelson1}.
One class of models stresses the coupling between electrons and 
intra-molecular phonons \cite{schluter}. 
Raman and infrared (IR) spectroscopy  have probed
the vibrational properties of C$_{60}$ compounds 
\cite{kratschmer,bethune,hare,dresselhaus1,martin1,kamaras,dresselhaus2} 
and many theoretical models have tried to explain
properties of the 46 distinct modes predicted by group theory.

The icosahedral ($\rm I_h$) symmetry of C$_{60}$  allows four distinct
IR active modes ($\rm T_{1u}$) and ten Raman active modes ($\rm 2A_{g} 
\bigoplus 8H_{g}$) in harmonic approximation.
It is customary to denote the IR modes at frequencies
528, 577, 1183, and 1429 \rm cm$^{-1}$,
as $\rm T_{1u}(i)$, i=1,2,3,4, respectively.
32 optically inactive (silent) modes are $\rm 1A_u, 3T_{1g}, 
4T_{2g}, 5T_{2u}, 6G_g, 6G_u$, and $\rm 7H_u$.
Higher order peaks are seen experimentally by increasing the optical 
depth of a sample. In principle
there are 380 second-order combination modes IR allowed by the 
$\rm I_h$ symmetry
\cite{martin1}. Second-order overtones are IR forbidden.

Several authors reported observation of weak modes in Raman
\cite{dresselhaus2,loosdrecht} and IR 
\cite{dresselhaus1,martin1,kamaras,chase} spectroscopy.
Wang et al. \cite{dresselhaus1}, Martin et al. \cite{martin1},
and Kamar\'{a}s et al. 
analyzed the weakly-active features in conjunction with Raman 
\cite{dresselhaus2} and neutron measurements \cite{neutron}
to extract the 32 fundamental frequencies of the silent modes.
The frequencies differ significantly among the authors
leaving the question of the assignment of fundamentals open.

Possible mechanisms of activating the weak modes
include $^{13}\rm C$ isotopic impurities, crystal environment effects
and anharmonicity. Impurities, dislocations and electric
field gradients at surface boundaries can be excluded due to their
sample dependence.
An experimental and theoretical vibrational study of 
$^{13}\rm C$-enriched crystals excluded the isotopic symmetry breaking as 
a potential candidate \cite{fabian}. 
A few of the weak modes
are thought to be activated due to the fcc crystal field effect.
The crystal field reduces the $\rm I_h$ symmetry of C$_{60}$   
and activates silent odd-parity modes. Above
260K the C$_{60}$ molecules freely rotate 
and the time averaged crystal field perturbation is zero.
This effect of `motional diminishing' of silent modes
have been experimentally observed and theoretically studied
by Mihaly and Martin \cite{martin2}. An experimental study
of pressure dependence of these modes would help to substantiate 
this mechanism.

The goal of the present paper is to identify, qualitatively,
the mechanism of activation of the higher-order vibrations;
detailed assignment to normal modes remains a task for the
future. Basic formalism of anharmonic effects on IR activity 
is given in \cite{herzberg,lax,szigeti,wallis,mills,sparks}.
There are two
ways in which anharmonicity can display itself in an optical 
spectrum.
Either it is driven by anharmonic interatomic forces 
(mechanical anharmonicity)
or by an anharmonic coupling of a photon field to two or more phonons
(electrical anharmonicity).
Although the two mechanisms are not independent,
 each has its own characteristic absorption intensity pattern. 
When compared with an experimental spectrum one can decide
which of the two kinds of anharmonicity prevails in the IR spectrum
of C$_{60}$. Although the spectrum may contain
cross-contributions from both phenomena, here they are treated
separately.

Several models have been used to calculate absorption
intensities in  harmonic approximation. Tight-binding models 
\cite{friedman,bertsch}
are in complete disagreement with the experimental results.
The bond-charge model
\cite{benedek} fits very well with frequency positions of 
fundamentals but the 
IR intensity pattern disagrees with basic trends in the observed spectrum.
The same is true for a Hubbard type model stressing electronic correlation 
effects \cite{kivelson}.
Relative intensities are best reproduced by the LDA approximation 
\cite{bertsch,giannozzi}. Due to its computational complexity the LDA scheme 
is not convenient for computing second order intensities.
We therefore propose a semi-empirical model which is satisfactory for
a qualitative comparison with experiment. Figure 1 summarizes the
performance of these models in calculating the absorption intensities.

Some characteristics of the experimental
IR spectrum \cite{martin1} are shown in Fig. 2. 
Combination (difference)
modes are higher-order modes with frequency $\omega$ equal
to $\omega_i \pm \omega_j$, the sum (difference) of fundamental
frequencies $\omega_i$. Their intensities are temperature 
dependent according to $ (n_i+\frac{1}{2})\pm (n_j+\frac{1}{2}) $, 
where $n_i$ is the Bose factor,
$n_i+\frac{1}{2} = \frac{1}{2} \coth (\frac{\hbar\omega_i}{2k_B T})$,
with a temperature $T$ and the Boltzmann constant $k_{B}$.
Following features can be observed in the spectra:
(i) besides four first-order peaks there are more than 180
weak absorptions,
(ii) no difference peaks are resolved (i.e. no temperature 
dependence of intensities except a trivial improvement in the frequency 
resolution at lower temperatures), (iii) most of
the spectral weight is in the high-frequency regime (1000 - 3000 \rm cm$^{-1}$),
and (iv) weak modes around four first-order bands are not enhanced through
a resonance effect. 
 
This paper treats the frequency positions and absorption
intensities independently.
Normal modes and frequencies
are calculated using a simple force-constant model proposed by Weeks
\cite{weeks}. This model fits IR data reasonably well but is not
expected to give especially realistic eigenfrequencies for the
silent modes. 
The dipole moment which arises due to the electron-phonon
coupling determines the absorption intensities \cite{wilson}.  
Only second-order combination and difference modes are considered 
in the paper.
Sec. \ref{sec:dyna} deals with the mechanical anharmonicity problem
with the Morse function used for the interatomic 
bond-stretching potential \cite{weeks}.  
A linear relation
between the dipole moment and ionic coordinates is proposed in this section.
The relation contains parameters fittable to the relative harmonic
absorption intensities.
Second-order modes are computed using a perturbation method 
ignoring possible resonances. However, the intensity pattern of the 
second-order modes 
fails to reproduce experimental features. 
An electrical anharmonicity model is therefore introduced in Sec. 
\ref{sec:elea}. 
Normal frequencies and normal modes are again taken to be those
 of the Weeks' model.
A semi-empirical model for an electronic
configuration on a distorted C$_{60}$ is presented which allows the 
electronic coordinates to depend in a nonlinear fashion on 
positions of ions. This gives
rise to an intensity pattern very similar to the experimental one.
Finally, conclusions are drawn in Sec. \ref{sec:conc}.

\section{MECHANICAL ANHARMONICITY MODEL}
\label{sec:dyna}

Considering the C$_{60}$ molecule as a system of oscillating ions with
electrons moving adiabatically in their field, the ionic dynamics
is governed by the potential:
\begin{equation} \label{eqn:potential}
V=\frac{1}{2}\sum\limits_{i=1}^{46}\sum\limits_{q=1}^{g_i}
m{\omega_i}^2Q_{iq}^2 + \frac{1}{6}\sum\limits_{i,j,k=1}^{46}
\sum\limits_{q,r,s=1}^{g_i,g_j,g_k} C_{iq,jr,ks}
 Q_{iq}Q_{jr}Q_{ks}.
\end{equation}
Here $Q_{iq}$ is the $q$-th normal mode coordinate belonging
to the frequency $\omega_i$, $i$=1,..,46, $q$=1,..,$g_i$, and
$g_i$ is the degeneracy of the $i$-th band. Higher-order terms
are neglected. The anharmonicity coefficients $C_{iq,jr,ks}$ are
given by
\begin{equation}
C_{iq,jr,ks}=\frac{\partial^3 V}{\partial Q_{iq}\partial Q_{jr}
\partial Q_{ks}}
\end{equation}
Light couples to the system via the term
\begin{equation} \label{eqn:coupling}
V_{1}=-{\bf \mu}(Q) \cdot \ {\bf E}, 
\end{equation}
where ${\bf E}$ is the externally applied macroscopic electric field 
and ${\bf \mu}$ 
stands for the dipole moment of the system. The latter is generally 
a nonlinear function of normal coordinates
\begin{equation} \label{eqn:dipole}
{\bf \mu} = \sum\limits_{i=1}^{46}\sum\limits_{q=1}^{g_i}
{\bf M}_{iq}Q_{iq} + \frac{1}{2}\sum\limits_{j,k=1}^{46}
\sum\limits_{r,s=1}^{g_j,g_k} {\bf M}_{jr,ks} Q_{jr}Q_{ks}
\end{equation}
Again, higher-order terms are not included and the folowing
formulas determine the expansion parameters ${\bf M}_{iq}$ and
${\bf M}_{jr,ks}$:
\begin{equation} \label{eqn:diff1}
{\bf M}_{iq} = \frac{\partial {\bf \mu}}{\partial Q_{iq}},
\end{equation}
and
\begin{equation} \label{eqn:diff2}
{\bf M}_{jr,ks}=\frac{\partial^2 {\bf \mu}}{
\partial Q_{jr} \partial Q_{ks}}.
\end{equation}
The vectors ${\bf M}_{iq}$ are nonzero only when the $Q_{iq}$ mode
is IR allowed.

Anharmonic dynamics ($C_{iq,jr,ks}\neq0$) and a linear coupling
of light to phonons (${\bf M}_{kr,ls}=0$) characterize the
mechanical anharmonicity (MA) phenomenon. 

Several force-constant models  for C$_{60}$ have been presented
\cite{weeks,force1,force2}. To calculate normal coordinates
 and the anharmonicity coefficients $C_{iq,jr,ks}$, I use the model suggested 
by Weeks \cite{weeks} which is a refined model of Weeks and Harter
\cite{harter}. This model contains two parameters which were fitted 
to selected IR and Raman frequencies. 
Ionic dynamics is governed by two types of interactions: 
(i) the Morse potential producing anharmonic terms 
 \begin{equation} V_m = \sum\limits_{i=1}^{90} D\{1-exp[-\alpha(r_i-r_{eq})]\}^2,
\end{equation} 
controls bond-stretching. Here D, $\alpha$, r$_{eq}$, and r$_i$ are,
respectively,
the dissociation energy, Morse anharmonicity, equilibrium and 
instantaneous length of the
$i$th bond. Summation runs over all bonds. The dissociation
energy is estimated as the average of the dissociation energies
of a single and a double C$_2$ bond, $D=5.0 \rm eV$, the
equilibrium length is taken to be $1.4{\rm \AA}$  and the parameter
$\alpha$ was fitted to the value $1.6 {\rm \AA}^{-1}$.
(ii). The bond-bending harmonic potential is given by
\begin{equation}
V_b = \sum\limits_{j}  \eta (\theta_{eq}-\theta _j)^2,
\end{equation}
where the summation is over the 60 pentagonal angles with the equilibrium 
angle of
$\frac{3}{5}\pi$ and the 120 hexagonal angles with the equilibrium angle of
 $\frac{2}{3}\pi$.
The potential does not distinguish between hexagonal and 
pentagonal angles and the best fit yields  $\eta = 12.48 \rm eV/rad$.

The bond-stretching potential in the harmonic approximation together with
the bond-bending potential give normal coordinates and
frequencies. The coefficients $C_{iq,jr,ks}$ come from the 
expansion of the Morse function to the third order in ionic 
distortions from equilibrium and from the transformation of the
Cartesian coordinates to the normal mode ones computed numerically.
Qualitative behavior of the normal modes of the model (with the 
bond-stretching potential in the  harmonic approximation) is discussed
in the original papers \cite{weeks,harter}.
It is enough to note that lower frequency normal modes exhibit mostly radial 
distortions while
the motion of higher-frequency ones  is tangential.

IR intensity of a given mode is proportional to the square of a 
dipole moment associated with the mode.
If ionic charges of the same value were put on the vertices of C$_{60}$, 
the resulting
dipole moment would be zero due to the center-of-mass conservation. 
The dipole activity
is therefore caused by changes in the electronic configuration. 
Carbon valence electrons fall into two classes. 
The first class  consists of
$\sigma$-electrons positioned
with the highest probability in the middle of bonds. These electrons 
have fixed
charges and do not contribute to the dipole moment (due to the 
center-of-mass conservation).
In the following the notion of a bond charge will include also a 
contribution
from ions in some effective way. The sign of such an effective bond 
charge will not
be important, it can be either positive or negative.
Allowing the bond charges to acquire a charge with 
dependence on the bond
lengths or by some other mechanism leads to a spectrum where 
the $\rm T_{1u}(2)$ mode is hardly visible instead of having 
the second largest activity \cite{benedek,jaro}.
 The second class consists of $\pi$-electrons which create a dipole moment
 in the following way.
Consider these
$\pi$-electrons to be vertex electrons moving in the field of 
their parent ions. Let these 
electrons interact further only with the three nearest ions. 
The  positions of the $\pi$-electrons are modeled in the following way. 
Denote ${\bf r}_i$ the radius-vector of the 
$i$-th electron measured from the vertex $i$ with the position ${\bf R}_i$
and  ${\bf R}_j^{(i)}$, $j=1,2,3$, 
the nearest ions positions, respectively, seen from the center of C$_{60}$.
The direction of ${\bf r}_i$ is taken to be the direction of the normal 
vector ${\bf n}_i$ to the plane given by three nearest ions with 
a rescaled position of
the one making the double bond with the vertex. This condition,
\begin{equation} 
{\bf n}_i \cdot ( {\bf R}_1^{(i)}- {\bf R}_2^{(i)} )=
 {\bf n}_i \cdot  ( {\bf R}_1^{(i)} - c_1 {\bf R}_3^{(i)} ) = 0, 
\end{equation}
introduces a fitting parameter $c_1$,
effectively measuring the ratio of the double- and single-bond charge
(here the bond ${\bf R}_i-{\bf R}_3^{(i)}$ is the double-one).
Single bonds are bonds connecting a hexagon with a pentagon and double-bonds
are connecting two hexagons. 
When there is more charge on the double-bond than on the single-one,
the parameter $c_1$ is greater than unity.
If the bond charge is negative, the direction is out of the bucky-sphere
 and if it is positive, the direction is inwards.

Consider the distance d of the vertex ion to the plane given by 
its three nearest ionic neighbors 
(with the double-bond neighbor rescaled as explained above).
Denote as  d$_{eq}$
the distance for the equilibrium configuration. 
Let, for a moment, the effective bond charge be negative.
If a distortion of the ionic
positions occurs such that $d > d_{eq}$ the vertex electron will be pushed `out'
of the C$_{60}$ sphere and vice versa. If the net bond-charge is positive, the 
situation is inverse. This phenomenology reflects a Coulomb repulsion 
(attraction)
of the vertex electron by (to) adjacent bonds. When these bonds move closer
together the vertex electronic cloud is deformed such that the mean electronic
position will be as far (close) as possible from (to) the bonds. 
The effective rate of the
deformation will be the second free parameter $c_2$ 
(the same for each vertex due to symmetry).
The relation
between the electronic position and the distance between the vertex ion and the
plane given by its nearest neighbors can then be expressed as follows :
\begin{equation} {\bf r}_i = \{1+c_2[d_i(c_1)-d_{eq}(c_1)]\} {\bf n}_i(c_1), 
\end{equation} 
where the dependence on the parameter $c_1$ is indicated.
The dipole moment is then clearly
\begin{equation}  {\bf \mu} = \sum\limits_{i=1}^{60}[1+c_2(d_i-d_{eq})]{\bf n}_i.
 \end{equation}
The normalization in both formulas is not important for calculating relative
values.
The distances $d_i$ depend for small distortions linearly on normal
coordinates, so only the linear term is kept here because the mechanical
anharmonicity couples this linear displacement to two normal modes.

There are two natural parameters in this model, $c_1$ and $c_2$. 
In the harmonic approximation
the intensity of the $j$-th mode  is \cite{wilson}
\begin{equation} \label{eqn:int1}
I_{\omega_j}^{(1)}=\sum\limits_{q=1}^{g_j} 
{\bf M}_{jq}^2. 
\end{equation}
Experimentally obtained relative intensities are 1., 0.48, 0.45, 0.378 
for the modes $T_{1u}(1)$, $T_{1u}(2)$, $T_{1u}(3)$, $T_{1u}(4)$,
respectively \cite{martin3}.
The best fit to these intensities yields the values $c_1=1.59$
and $c_2=0.67{\rm \AA}^{-1}$.
The IR spectrum obtained with the fit (all peaks in this and following 
figures have the Lorentzian widths taken to be uniformly 2$\rm cm^{-1}$) 
along with an experimental one is shown in Fig. 3. 
Agreement with experiment is very good.

For the frequencies that are not in the immediate neighborhood of 
the frequencies of the four IR allowed fundamentals, the following
formulas were obtained in Ref. 18 for the second-order intensities 
of combination and difference modes: 
\begin{eqnarray}  \label{eqn:int2} 
I_{\omega_k+\omega_l}^{MA} =&\frac{\hbar}{2m^3} 
\frac{\omega_k+\omega_l}{\omega_k \omega_l} 
(1+n_k+n_l) \\ &\times \sum\limits_{r,s=1}^{g_k,g_l}
(\sum\limits_{j\in IR} \frac{\langle {\bf M}_j|C_{j,kr,ls} \rangle}
{{\omega _{j}}^{2}-(\omega_k+\omega_l)^2})^2, \nonumber
 \end{eqnarray}
and 
\begin{eqnarray} \label{eqn:int3}
I_{\omega_l-\omega_k}^{MA} =&\frac{\hbar}{2m^3} 
\frac{\omega_l-\omega_k}{\omega_k 
\omega_l} 
(n_k-n_l) \\& \times \sum\limits_{r,s=1}^{g_k,g_l}
(\sum\limits_{j\in IR} \frac{\langle {\bf M}_j|C_{j,kr,ls} \rangle}
{\omega^2_j-(\omega_l-\omega_k)^2})^2, \nonumber 
\end{eqnarray} 
respectively.
The summation in brackets is over four IR active bands
and the inner-product notation stands for the sum over a degenerate
set :
\begin{equation} 
\langle {\bf M}_j | C_{j,..} \rangle \equiv \sum\limits_{q=1}^{g_j}
{\bf M}_{jq} C_{jq,..}.
\end{equation}
When the frequency of a combination (difference) mode
is near the frequency of an IR allowed mode (the Fermi
resonance effect), a perturbation
leads to a mixing of the two
modes and spreads out their frequencies 
(see Ref. \cite{herzberg}). 
The second-order modes are enhanced conserving the original
spectral weight so the integrated absorption intensity of 
the band is unchanged by the anharmonic perturbation.  
If the spectral resolution is not enough to resolve the
two modes the resulting picture is similar to the original
one without a perturbation. The Fermi resonance effect has not been observed
in C$_{60}$.

Figure 4 shows the results of the numerical calculations based on
the Eqs. \ref{eqn:int1}-\ref{eqn:int3}. 
Some trends in the
spectrum are clear already from the equations. First of all the second-order
intensities are relatively  weak compared to the experimental
spectrum in Fig. 2 (the experimental picture here is somewhat misleading
due to the saturation of first-order peaks). Most intense modes have frequencies close
to the four IR bands, leaving  high-frequency combination
modes practically invisible. Moreover there are relatively
intense difference modes (identified by their strong temperature 
dependence) in the lower part of the spectrum. These features are 
in contradiction to experiment thus excluding mechanical
anharmonicity as the mechanism for activation of the combination
modes seen in experiment.
In matching the combination modes to experimental data, authors in
Ref. 8 did not find any evidence for a significant deviation of the
frequencies of these modes from the values of $\omega_i+\omega_j$.
This supports the above conclusion that mechanical anharmonicity
is not producing significant effects in the C$_{60}$ IR spectrum,
since the relative frequency shift as a consequence of mechanical
anharmonicity only is of the same order of magnitude 
as the relative intensities of the second-order modes.

\section{ELECTRICAL ANHARMONICITY MODEL}
\label{sec:elea}

The electrical anharmonicity (EA) is a less studied phenomenon of molecular
physics than the mechanical one. It is based on the fact that the dipole 
moment is generally a nonlinear function of normal modes. 
In view of Eqns. \ref{eqn:potential} and \ref{eqn:dipole}, electrical
anharmonicity arises from the second term in Eq. \ref{eqn:dipole}, 
while the ionic dynamics is harmonic ($C_{iq,jr,ks}$=0). 
Selection rules for the second-order modes are reflected in the elements of 
the matrix ${\bf M}_{jr,ks}$, and
are the same as in the case of the mechanical anharmonicity. Since the 
ionic dipole moment is linear in ionic positions it is clear that the nonlinear
contribution stems from a nonlinear response of electronic positions 
to a change in ionic configuration. A harmonic treatment now suffices for the
ionic displacements; the Weeks model of Sec. II. is used.

The nonlinear electronic response is modeled in the following way. 
The notation is the same  as in the previous section. 
Consider again a $\pi$-electron in the field of its parent ion and
adjacent bond-charges.
The interaction with its nearest neighbor ions is governed by the Coulomb potential
\begin{equation}
V_{e-i}({\bf r}_i)=-\kappa_{i} \sum\limits_{j} \frac{1}{|{\bf R}_i -
{\bf R}_j+{\bf r}_i|} 
\end{equation}
and similarly the interaction with adjacent bond-electrons is given by 
\begin{equation} 
V_{e-be}({\bf r}_i)=\kappa_{be} \sum\limits_{j} \frac{1}{|\frac{{\bf R}_i
 -{\bf R}_j}{2}+{\bf r}_i|}
\end{equation}
Summations are over the three nearest ions and ${\bf R}_i$ is the position 
of the vertex ion. 
Note that while ${\bf R}$'s are measured from the mass center of C$_{60}$,
 ${\bf r}_i$ is measured from the position of the $i$th vertex ion
 (${\bf R}_i$).
The strengths of the interactions are measured by some effective
charges $\kappa_i$ and $\kappa_{be}$ for neighbor ions and adjacent 
bond-electrons respectively.
Only the ratio $\frac{\kappa_i}{\kappa_{be}}$ is a relevant fitting parameter.
The motion of the $\pi$-electron in the field of its vertex ion is simplified by
restricting it to a sphere around the ion with a radius $R$ which will
be the second fitting parameter:
\begin{equation} {\bf r}_i = R {\bf n}_i. 
\end{equation} 
This gives a simple two-dimensional minimization scheme: for each vertex and
  a pair of fitting parameters
$(R,\kappa_i/\kappa_{be})$ find a unit vector ${\bf n}_i$ such that the function 
\begin{equation} \label{eqn:potential1}
V_{e-i}({\bf n}_i) + V_{e-be}({\bf n}_i)
\end{equation}
is minimal. The electrical dipole moment is then computed  and resulting
 first-order intensities (Eq. \ref{eqn:int1}) compared with corresponding 
experimental values. The best fit corresponds to  values
of $R=0.06{\rm \AA}$ and $\kappa_i/\kappa_{be}=4.80$. For some range of the 
parameters
there are two electron positions for which the potential in Eq. \ref{eqn:potential1}
has a local minimum.
In such cases the global one was considered. The best fit lies in 
the region with one minimum. It is obvious that the best fits have no physical 
justification. 
To support the model I did simulations with different, more physical values of
the free parameters obtaining the same qualitative picture as will be shown later.
It is also appropriate to remark that a feedback from the adiabatic changes in 
electronic positions to ionic motion is implicitly considered
in the harmonic level in the force-constant model.

For the IR absorption the changes of the minima positions with ionic 
distortions are relevant. Numerical differentiation was used to obtain 
the dipole-moment matrices
${\bf M}_{iq}$ and ${\bf M}_{jr,ks}$  from Eqns.
\ref{eqn:diff1} and \ref{eqn:diff2}.
An important feature of the model is that the $\pi$-electronic positions are 
more sensitive to tangential distortions than to radial ones.       
 
Second-order absorption intensities of combination and difference modes 
have now simple forms \cite{szigeti}:
\begin{equation} \label{eqn:int5} 
I_{\omega_k+\omega_l}^{EA} = \frac{\hbar}{2m} 
\frac{\omega_k+\omega_l}{\omega_k \omega_l} 
(1+n_k +n_l )
\sum\limits_{r,s=1}^{g_k,g_l} {\bf M}_{kr,ls}^2 
\end{equation}
\begin{equation} \label{eqn:int6}
I_{\omega_k-\omega_l}^{EA} = \frac{\hbar}{2m} 
\frac{\omega_k-\omega_l}{\omega_k \omega_l} 
(n_l - n_k ) 
\sum\limits_{r,s=1}^{g_k,g_l} {\bf M}_{kr,ls}^2 
\end{equation}
Figure 5 shows the spectrum obtained from equations \ref{eqn:int5} and
\ref{eqn:int6}. The following features can
be extracted. Overall intensity of the weak modes is higher (in relative
sense) than in the case of the mechanical anharmonicity. Spectral weight is
shifted towards higher frequencies. This is a consequence of high sensitivity of
electronic positions to tangential distortions which are 
characteristic for higher-frequency modes. The sensitivity of 
electrons to the tangential ionic motion is also the reason why difference
peaks have relatively very small intensity (the difference peaks are most 
intense in the region 
of 600 - 1000 $\rm cm^{-1}$, however the intensities are much smaller than those of
combination modes in the region 1000 - 3500 $\rm cm^{-1}$). There is obviously no
resonance effect since the two terms in the Eq. \ref{eqn:dipole} are independent.
The frequency distribution in the Weeks model differs from that in C$_{60}$ so a 
closer comparison with experiment is not possible. One consequence is that 
in Figure 5 
weak features up to $4000\rm cm^{-1}$ are visible, while experimentally 
weak peaks above
$3500 \rm cm^{-1}$ have not been resolved. This difference in the frequency
distribution may be a part of the reason why
there is  so little activity in the region $600-1000 \rm cm^{-1}$. Note
that almost all of the peaks experimentally observed in this region were associated
with modes IR forbidden in the second-order \cite{dresselhaus1,martin1} and their 
appearance must be accounted
for by other mechanisms.

\section{CONCLUSION}
\label{sec:conc}
Mechanical and electrical anharmonicity provide possible
mechanisms for activating weak modes resolved in IR spectra of
C$_{\rm 60}$ thin films and single crystals. I have proposed simple 
semi-empirical models of the
phonomena. The main features of the models are: (i) separation of 
ionic dynamics and mechanism of optical activation (the models can
be used for any set of normal modes), and (ii) emphasis on $\pi$-electronic
system rather than on bond charges. Both models give a spectrum of 
combination and difference
modes which is compared with IR measurements. It is found that 
mechanical anharmonicity exhibits features different than those
observed. These features can be generally expected from basic formulas
(e.g. those of Eqs. \ref{eqn:int2} and \ref{eqn:int3}) and the model
described in Sec. II
only helps to visualize them. As a by-product the intensities 
of four first-order IR allowed bands are well reproduced.

The electrical anharmonicity model introduced in Sec. III is based on 
a nonlinear response of 
$\pi$-electronic configuration to ionic distortions. Now the
absorption spectrum has fewer characteristics given a priori by a
theoretical formula and is more model-dependent. The main feature the
model which leads to quite successful comparison of its spectrum
with experiment is that electronic positions are much more 
sensitive to tangential ionic motions than to radial ones.          

The separation of mechanical and electrical anharmonicity is
posteriorly justified by the dominance of the latter. However, 
the IR activity around four first-order
peaks is caused by mechanical anharmonicity due to resonance effects,
as discussed in Sec. II.
There is still a region of optical activity  ($\rm 600-1000 \rm \rm  cm^{-1}$)
which this simple model cannot explain. Although trial assignments
exclude most of the observed peaks in the region as combination modes,
the question is still an open one and more sophisticated quantum-mechanical 
treatment can yield more authoritative results.

\section{ACKNOWLEDGEMENTS}

I am grateful to P. B. Allen for proposing this study.
I thank M. C. Martin and L. Mihaly for useful discussions.
This work was supported by NSF Grant No. DMR 9417755

\newpage

{\bf FIGURES }

FIG.~1. Comparison of calculated relative absorption intensities of
IR allowed $\rm T_{1u}(i)$, i=1, 2, 3, and 4,  modes with experiment.
Intensities of the band $\rm T_{1u}(1)$ are taken to be unity.

\vspace{2em}
FIG.~2. C$_{60}$ single crystal IR transmission spectra at  300K and 77K by M. Martin
et. al. \cite{martin1}.

\vspace{2em}
FIG.~3. First-order IR allowed intensities calculated in Sec. II and experimentally
obtained spectrum (inset) by  Hare et.al \cite{hare}.

\vspace{2em}
FIG.~4. IR spectra at 300K and 77K computed using the mechanical anharmonicity 
model introduced in Sec. II.
Difference modes are easily identified by their strong temperature dependence, while
combination modes show no such trends. 

\vspace{2em}
FIG.~5. The electronic anharmonicity model (Sec. III) produces absorption spectra
which show similar trends as experimental ones. Difference peaks carry very little
spectral weight comparing to high-frequency combination ones.

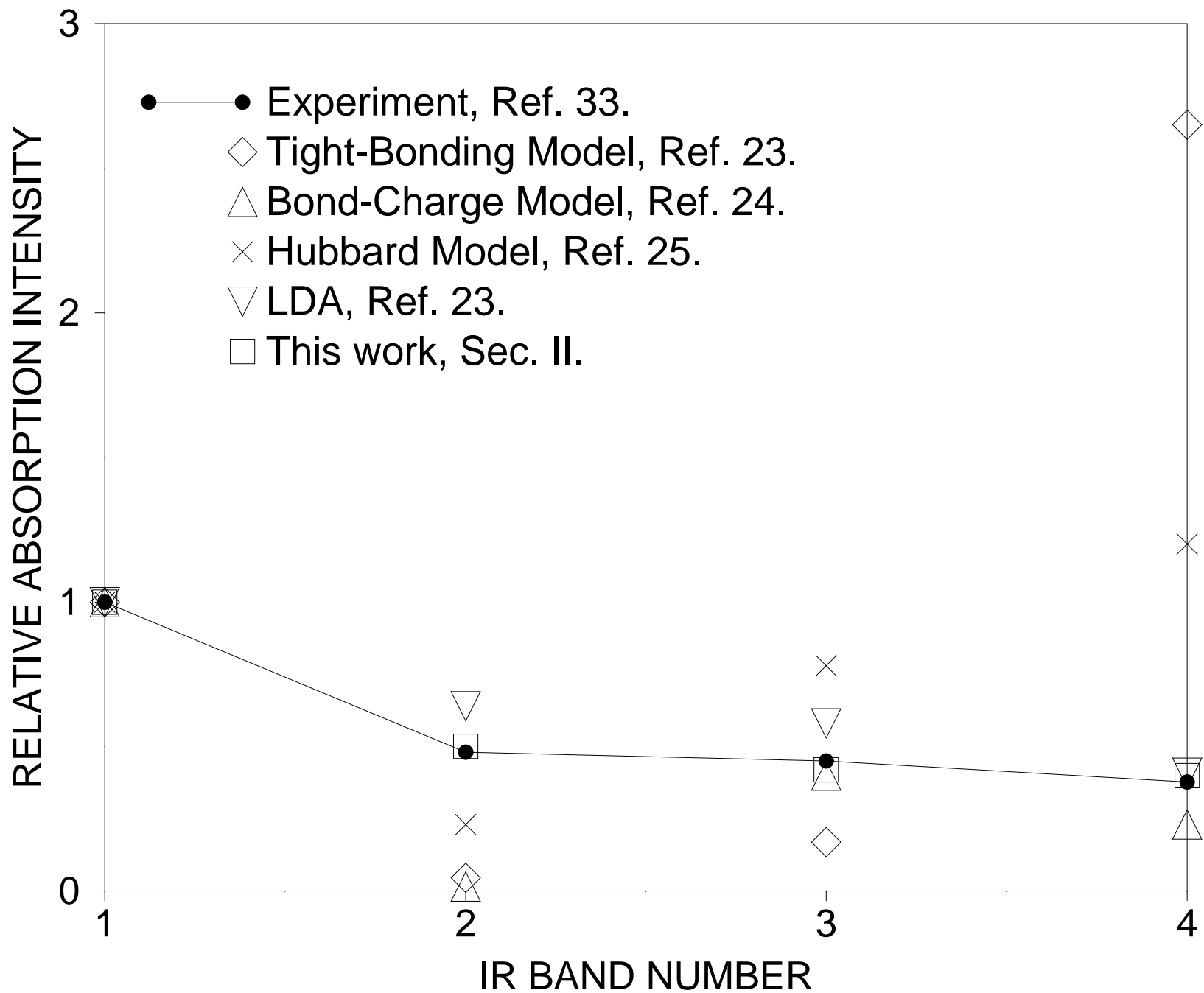
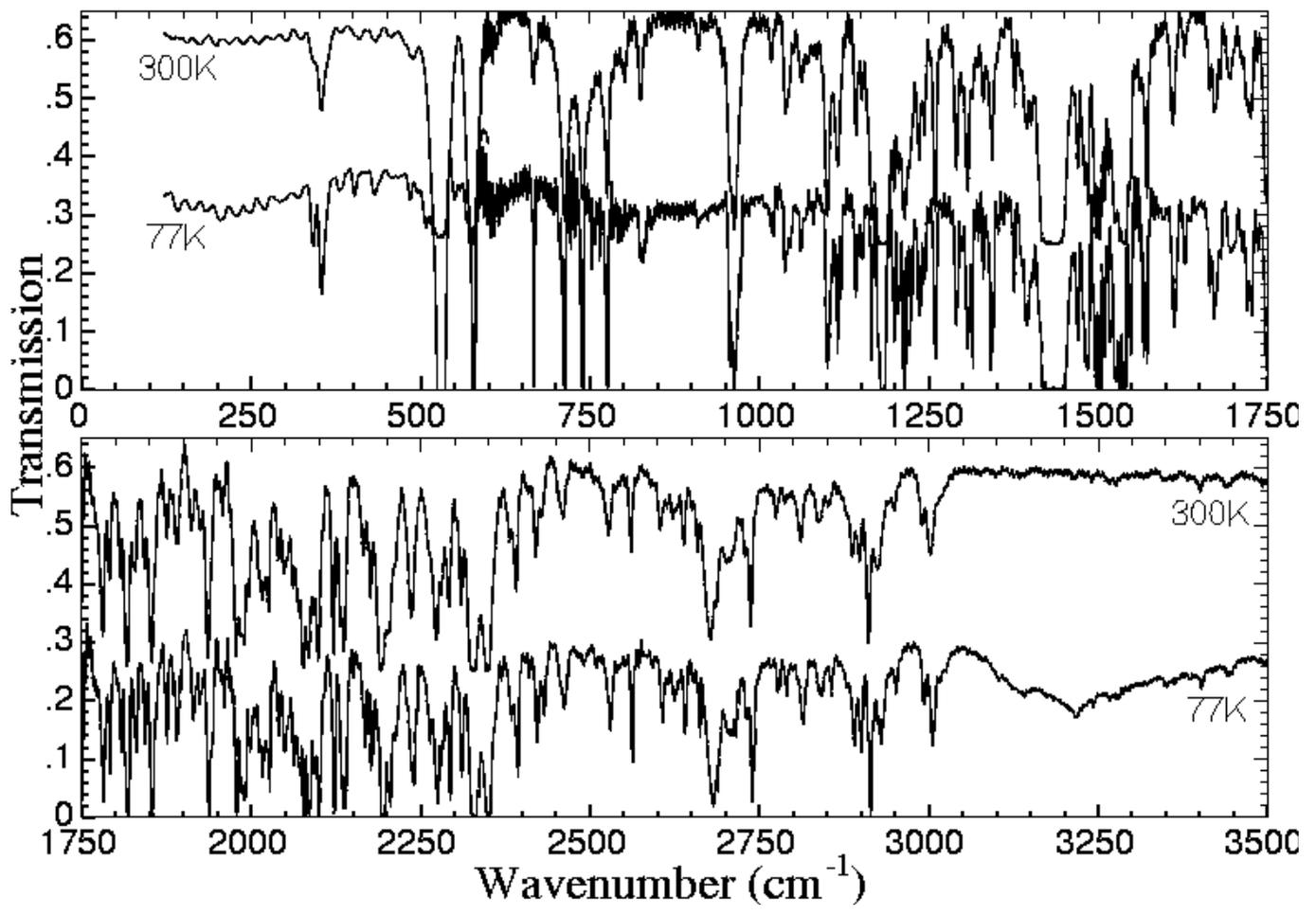
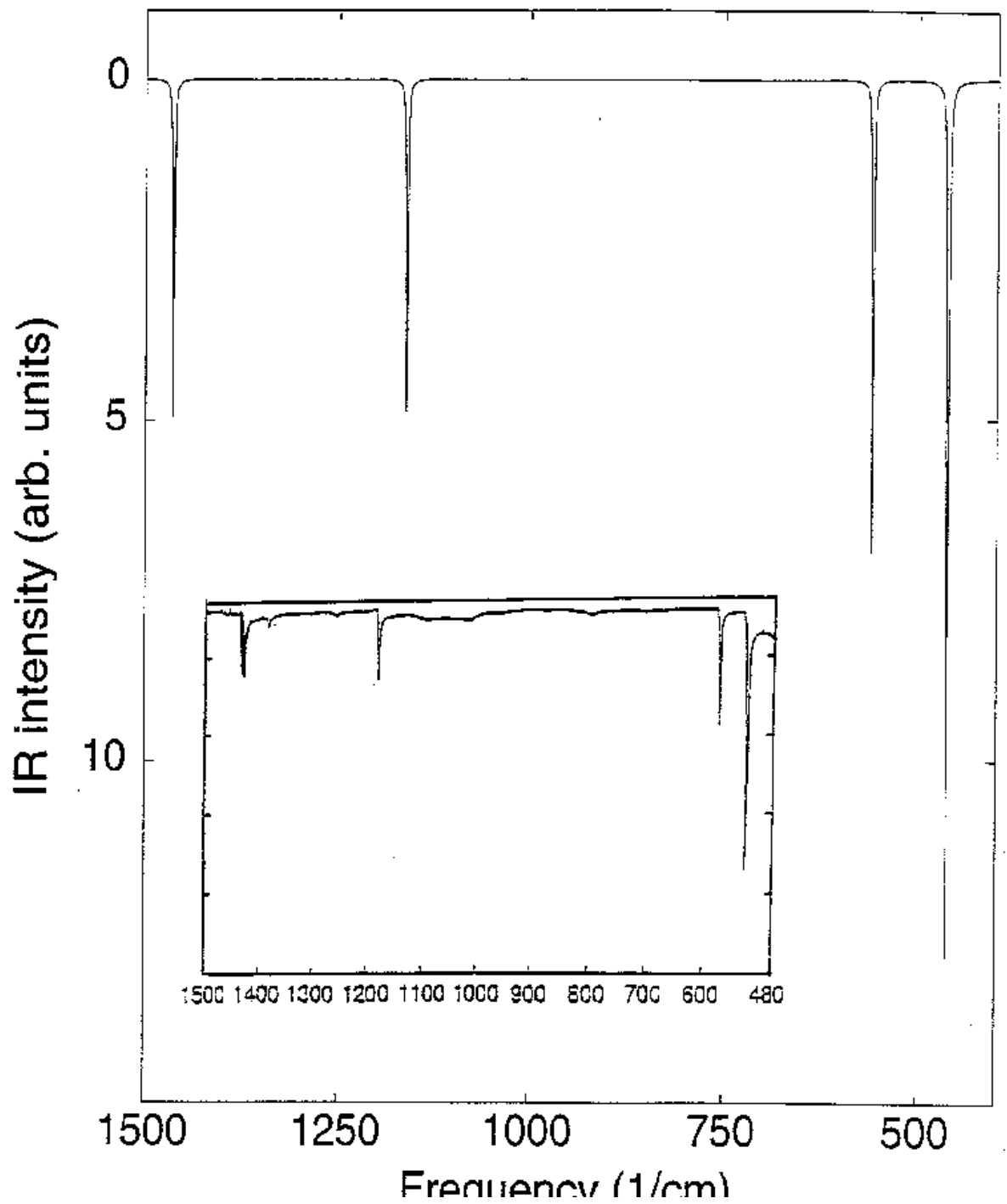
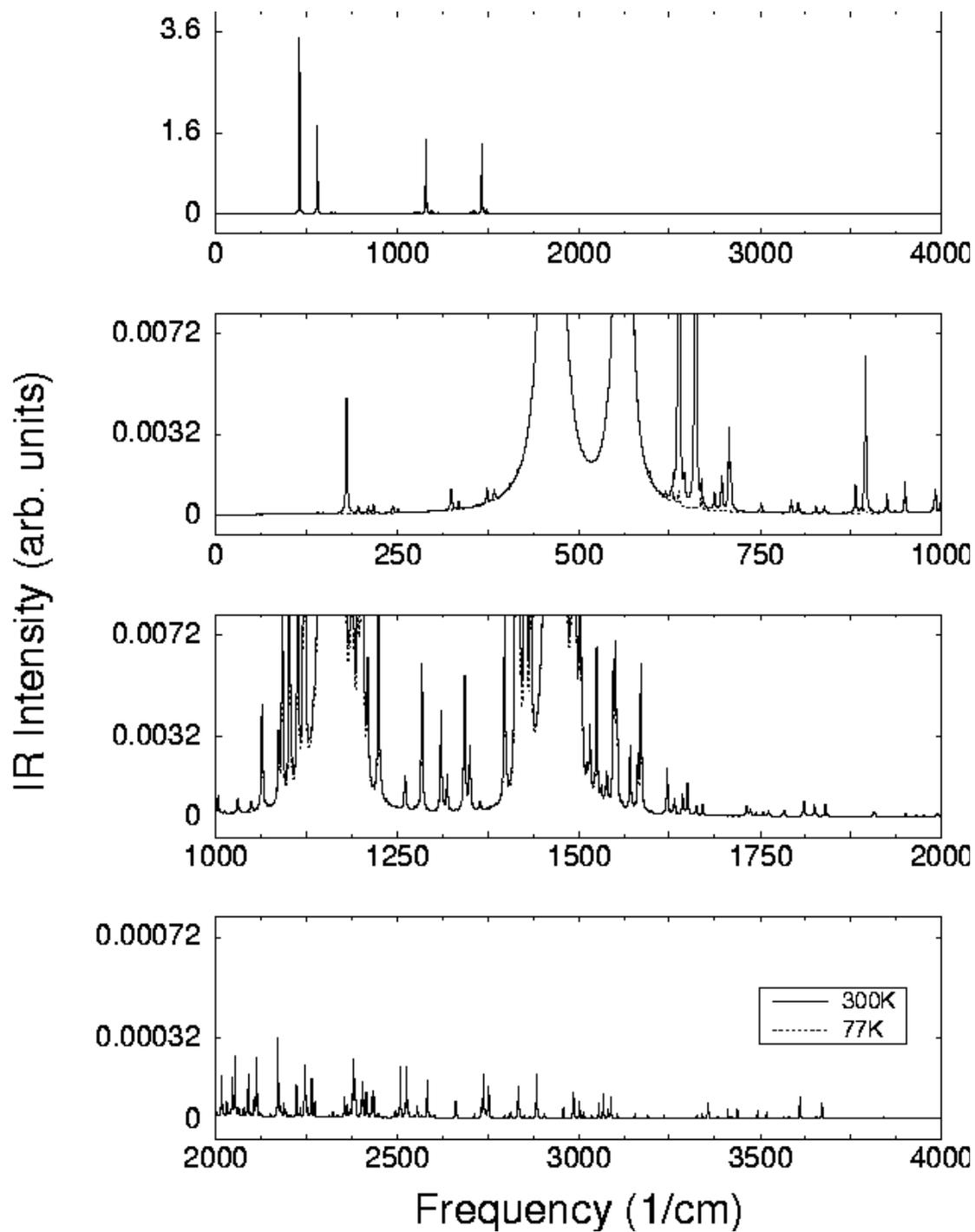
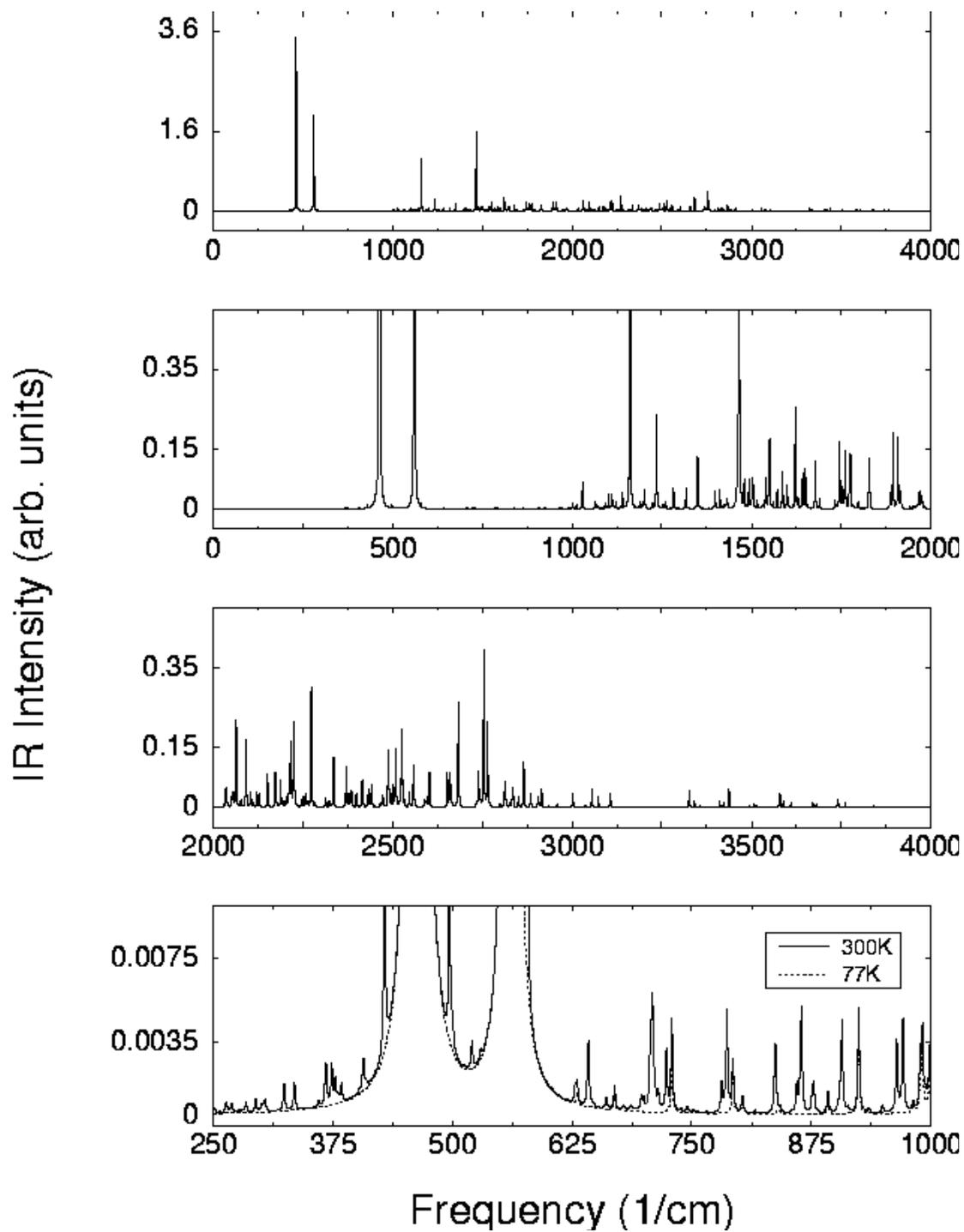
\end{document}